\documentclass[sigconf]{acmart}

\usepackage{tikz}
\usepackage{makecell}
\usepackage{booktabs}
\usepackage[htt]{hyphenat}  

\newcommand{\olgaDONE}[1]{
}
\newcommand{\olgaDone}[1]{
}

\AtBeginDocument{%
  \providecommand\BibTeX{{%
    \normalfont B\kern-0.5em{\scshape i\kern-0.25em b}\kern-0.8em\TeX}}}

\begin{document}

\title{Predicting Engagement with Sponsored Content \\ Across Account Types on Instagram}

\author{Pedro~Victor~de~Sousa~Lima~and~Olga~Goussevskaia}
\email{pvsousalima@gmail.com, olga@dcc.ufmg.br}



\renewcommand{\shortauthors}{TBD}

\begin{abstract}
In this work we explore how users interact with sponsored content on social media platforms by collecting and analyzing a large-scale dataset of sponsored Instagram posts. To maintain transparency, we favor robust statistical analysis and explainable models over deep learning techniques. Our pipeline categorizes Instagram accounts along multiple dimensions, including audience size and entity type. We complement this with semantic features extracted from post captions and hashtags, and use these elements to train regression models that forecast engagement.

We validate our approach on a dataset comprising over 15M Instagram posts authored by over 700K accounts featuring sponsored content. 
Our analysis shows that per-post engagement is partly predictable and highly optimized when accounts are segmented by entity type or audience tier.
Our models achieve competitive predictive power,
while remaining fully transparent about which features drive engagement.

\end{abstract}


\maketitle

\section{Introduction}
\label{sec:intro}
\olgaDONE{TODO REVISE, CHECK, AI-help:}
Sponsored content has become one of the primary monetization mechanisms on
online social networks, with influencers and brands routinely partnering to
promote products through ordinary-looking posts. As this form of
advertising has grown, so has the need to understand what actually drives
its performance: which accounts benefit most from sponsored partnerships,
which posting strategies translate into audience engagement, and whether
these dynamics are consistent across the highly heterogeneous population of
accounts that participate in influencer marketing, ranging from individual
content creators to brands, products, and organizations. Answering these
questions matters not only to platforms and advertisers seeking to design
effective campaigns, but also to researchers and regulators interested in
understanding the broader effects of monetized content on user behavior and
on the integrity of social media ecosystems.

A central challenge in this space is that engagement with sponsored content
is not a uniform phenomenon. The same posting pattern, for instance, the use
of a particular hashtag, a specific posting frequency, or a given tone in
the caption, may resonate very differently depending on who is posting and
to whom. An individual influencer with a small, highly engaged following
may benefit from cues that would have little effect on a brand account with
millions of followers, and vice versa. Treating all sponsored posts as
draws from a single, homogeneous population, as much of the existing
engagement-prediction literature implicitly does, risks averaging away
exactly the heterogeneity that matters most for understanding monetization
performance in practice.

Research on sponsored content engagement has progressed along two largely
separate tracks. The first focuses on detecting and characterizing
sponsored content itself, treating disclosure compliance or the aggregate
evolution of advertising activity as the primary object of study
\citep{kim2021discovering,martins2022characterizing,bertaglia2024instasynth}.
The second, more directly relevant to our goals, treats post-level
engagement as a prediction target. The current state of the art in this
track favors classification or regression models trained on multimodal post
features, image, caption text, and influencer-level metadata, with interpretability often prioritized over raw predictive accuracy so that the resulting feature importances can be translated into actionable posting guidelines~\citep{tricomi2023follow,hsiao2025decoding}.

These approaches
have demonstrated that simple, interpretable models can match or exceed
deep-learning baselines while remaining transparent about which features
drive engagement, and they have shown that influencer tier and content
category are both relevant moderators of engagement outcomes. A smaller
body of work zooms in on a single influencer category or a single
cross-platform behavior to study engagement determinants in more depth
\citep{cruz2024characterizing,bouchardet2025call2go}, while longitudinal
studies of disclosure practices provide complementary evidence that
monetization outcomes differ substantially between influencer tiers over
time~\citep{bertaglia2025disclosure}.

What is missing from this picture is a systematic account of how the
\emph{same} engagement-prediction model behaves once the account population
is explicitly partitioned along the dimensions that plausibly drive this
heterogeneity, namely entity type (individual, brand, product, geographic
location, or other) and follower-base size, jointly. Prior engagement
prediction work segments accounts by topical category or by audience tier
in isolation~\citep{tricomi2023follow,hsiao2025decoding}, but does not
distinguish accounts by what they \emph{are}, nor does it
test whether feature importance itself, rather than just raw engagement
levels, varies across these segments. Conversely, the studies that do
report differences in engagement and monetization across influencer types
or platforms~\citep{bertaglia2025disclosure,bouchardet2025call2go} are
observational analyses of aggregate trends rather than predictive models
that quantify which specific posting-pattern features matter, and by how
much, within each segment. Our work closes this gap. We collect and
analyze a large-scale dataset of Instagram posts containing sponsored
partnerships, partition accounts by entity type and follower-base size, and
incorporate semantic analysis of post text and hashtags as core predictive
features. In summary, our contributions are threefold:

\begin{itemize}
    \item We frame engagement prediction for sponsored content as a
    regression problem---predicting a post's likes and comments---trained
    separately across segments defined jointly by account entity type and
    follower-base size, rather than treating the influencer population as
    homogeneous or segmenting along a single dimension at a time.
    \item We incorporate semantic analysis of textual content and hashtags
    as first-class predictive features, alongside the more commonly used
    structural and influencer-level metadata, and estimate the relative
    importance of these features within each account segment.
    \item We show that the same posting pattern can have a markedly
    different, and at times opposite, impact on engagement performance
    depending on an account's entity type and follower-base size, a
    finding that prior aggregate-level studies suggest but do not directly
    quantify, and which has practical implications for how influencers and
    brands should tailor their ad placement strategy.
\end{itemize}

We validate our approach on a dataset comprising over 15M Instagram posts authored by over 700K accounts featuring sponsored content. 
Using interpretable, tree-based models, rather than deep architectures, we find that a meaningful share of engagement is predictable from the post alone: a baseline model reaches a coefficient of determination $R^2{=}0.40$ for likes and $R^2{=}0.49$ for comments.
Our models achieve competitive predictive power, reaching up to $R^2=0.74$ for the most predictable entity types. We show that content features dominate prediction; entity classification serves as the strongest structural complement, while explicit topics are largely redundant. The structural advantage of our entity-based lens stems from its capacity to isolate institutional profiles whose engagement dynamics operate under fundamentally different rules than individual creators. Among individual creators, we show that mid-tier influencers are the most content-predictable, and top-tier accounts the least.



The rest of the paper is organized as follows.
Section~\ref{sec:related} reviews related work;
Section~\ref{sec:data_engineering} describes data collection, account classification,
topic modeling, and our metrics; Section~\ref{sec:data_characterization} characterizes the
corpus across entity types and follower tiers; Section~\ref{sec:results}
presents the engagement-prediction models, their per-entity performance, and the
feature ablation; and Section~\ref{sec:conclusion} concludes the paper.

\section{Related Work}
\label{sec:related}

\olgaDONE{TODO include citation to our work (Lucas, Emanuelle}
\olgaDONE{TODO include citation to Webmedia 2025 (OK $\times 1$), 2024, 2023, \ldots}
\olgaDONE{TODO REVISE, CHECK Text Flow, AI-help}

Research on sponsored content in Online Social Networks (OSNs) has developed along several lines, such as: detecting and characterizing sponsored posts, predicting user engagement, predicting user engagement with interpretable models, and analyzing how engagement and monetization strategies vary across influencer types and platforms. We review each line below and position our contribution relative to it.

\textbf{Detecting and Characterizing Sponsored Content.}
A substantial body of work focuses on identifying sponsored posts,
particularly those that fail to disclose sponsorship as required by
advertising regulations. \citet{kim2021discovering} propose a
learning-to-rank model that detects undisclosed sponsorship by jointly
learning textual, visual, and social-network aspects of a post, using a
heterogeneous graph over influencers, brands, and posts. More recently,
\citet{bertaglia2024instasynth} explore whether large language models can
generate realistic synthetic Instagram captions to augment training data for
this same detection task, evaluating the trade-off between the fidelity of
generated captions and their downstream utility for classification.
\citet{martins2022characterizing} take a more descriptive approach,
characterizing the evolution of sponsored content on Facebook and Instagram
across the COVID-19 pandemic using a cross-platform dataset of\olgaDONE{TODO CHECK:} 9.5 million
sponsored posts, with a particular focus on how engagement with sponsored
content varied by topic over time. These works establish the methodological
foundation for working with large-scale sponsored-post datasets, but their
central goal is to determine how
sponsorship patterns evolve in aggregate, rather than to model what drives engagement. 
\citet{olivetti2024through} investigate content trends on Instagram by categorizing visual media through automated metadata.

\textbf{Engagement and Popularity Prediction.}
Prediction algorithms for user engagement have received a lot of attention, ranging from classic linear models (Logistic Regression and Support Vector Machines) \cite{talamadupula2015predicting} to (temporal) cascade-centric models \cite{aravamudan2023anytime}, State Space Models \cite{iccmamba2025}, multimodal (behavioral) feature integration \cite{emotionalengagement2025, li2025vquala}, and deep graph architectures and transformer networks \cite{wwwcompanion2024}. 

\textbf{Prediction with Interpretable Models.}
\citet{tricomi2023follow} build classifiers to predict Instagram post
engagement over a dataset of\olgaDONE{TODO CHECK:} 10 million posts from 34 thousand influencers,
explicitly favoring interpretable, statistical models over deep
learning so that the resulting feature importances can be turned into
actionable posting guidelines; they show that influencer tier (audience
size) and content category are both crucial to engagement outcomes.
\citet{hsiao2025decoding} pursue a related goal using a different feature
decomposition, combining image, image-topic, text-topic, and influencer
features to predict the popularity of sponsored posts specifically, and find
that no single feature type is sufficient on its own. At a narrower scope,
\citet{cruz2024characterizing} focus on a single influencer category
(fashion), correlating engagement metrics with post metadata such as caption
sentiment, length, and hashtag usage. Our work adopts the same general
strategy, classification-based engagement prediction paired with feature
importance analysis, but differs in scope: rather than focusing on a single
category or treating account type as one feature among many, we explicitly
partition accounts by entity type and follower-base size \emph{before}
modeling, allowing us to compare how the same posting pattern's importance
shifts across segments rather than estimating a single global importance
ranking. 
\citet{prenassi2025narrativas} utilizes large-scale thematic parsing and explicit linguistic feature extraction to analyze how the text configurations adopted by digital influencers affect audience attention and drive distinct engagement loops.

\textbf{Cross-Platform and Influencer-Type Effects.}
A third line of work asks how engagement and monetization dynamics differ
across influencer types and platforms. \citet{bertaglia2025disclosure}
conduct a multi-country longitudinal study of disclosure practices and find
that mega-influencers sustain consistently high engagement regardless of
disclosure, while micro-influencers struggle to achieve above-average
engagement specifically on sponsored posts, indicating that monetization
performance is far from uniform across the influencer population.
\citet{bouchardet2025call2go} similarly find that, in a cross-platform
setting, child and adult influencers exhibit different relationships
between follower count and engagement, with child influencers growing their
audiences faster while adult influencers sustain proportionally higher
engagement. Both studies support the premise underlying our work: that
follower-base size and the nature of the account meaningfully change how
posting behavior translates into audience engagement.

\textbf{Comparison to Our Work and Contribution.}
Taken together, this body of work establishes that (i) sponsored content can
be reliably detected and characterized at scale, (ii) post-level engagement
can be predicted from textual, visual, and influencer-level features using
interpretable models, and (iii) engagement and monetization dynamics differ
across influencer tiers and, to a more limited extent, across platforms.
However, prior engagement-prediction studies, including the closest
comparison points, \citet{tricomi2023follow} and
\citet{hsiao2025decoding}, treat the influencer population largely as a
single pool segmented at most by topical category or audience tier,
and do not explicitly distinguish accounts by \emph{entity type}, that is,
whether an account represents an individual, a brand, a product, a
geographical location, or another kind of entity. Similarly, the studies
that do report follower-base or influencer-type effects on engagement
\citep{bertaglia2025disclosure,bouchardet2025call2go} are observational
analyses of aggregate trends rather than predictive models that quantify
feature importance conditioned on these account properties. Our main
contribution is to combine these two perspectives: we model engagement
prediction as a classification problem, as in \citet{tricomi2023follow}, but
we partition accounts jointly by entity type and follower-base size prior to
modeling, and we further incorporate semantic analysis of textual content
and hashtags as first-class features rather than auxiliary metadata. 

\section{Data Engineering}
\label{sec:data_engineering}

Our methodology comprises four stages. We first collect a large corpus of sponsored Instagram posts, using sponsorship-disclosure hashtags as filters~\cite{martins2022characterizing}. We then classify each posting account along two axes---the size of its follower base (micro/macro/mega tiers) and its entity type (a six-way split). We next model latent themes from hashtag co-occurrence. Finally, we define the engagement metrics that serve as prediction targets: the number of interactions (likes and comments) with a post, and its engagement rate (interactions normalized by the size of the respective follower base).

\subsection{Data Collection}
\label{sec:collection}

Data were collected via \texttt{CrowdTangle}, a public \texttt{Meta analytics} platform that supported monitoring of public content across Facebook, Instagram, and Reddit~\cite{crowdtangle_about_us}. Given the limited access of Instagram's official API to third-party timelines, \texttt{CrowdTangle} was the only viable option for large-scale collection; the platform was discontinued in 2024, limiting our dataset to that year.

 We built three complementary datasets using sponsorship-dis\-clo\-sure hashtags (e.g., \texttt{\#advertise\-ment}, \texttt{\#spon\-sored}, \texttt{\#ad}, \texttt{\#paid},
  \texttt{\#pro\-motion}) as collection filters: (i) a historical, hashtag-filtered corpus of $\approx$15.5M posts~\cite{martins2022characterizing}; (ii) the most recent posts from the highest-interaction accounts appearing in those posts; and (iii) the most recent posts from the most-mentioned accounts (i.e., likely sponsor profiles). The latter two datasets were further expanded by retrieving 100-post histories for top-mentioned/highest-interacted-with accounts, supporting our sponsor-side analysis. Together, the three datasets comprise $\approx$17.6M collected posts.

\subsection{Preprocessing}
  Posts were filtered to remove noise, including non-Latin encodings. Separately, for topic modeling, hashtags used by fewer than 250 unique users were discarded. After
  classification, the working corpus comprises $\approx$\textbf{15.7M} posts authored by $\approx$\textbf{766K} classified accounts, distributed across the six entity types
  detailed below (Table~\ref{tab:entities}).

\subsection{Accounts Classification}
We classify accounts along two complementary axes: the size of their follower base and the type of entity each account represents. The first captures reach---how large an audience a post can plausibly attract---while the second captures \emph{what kind} of actor publishes the content, distinguishing individual creators from brands, products, places, events, and creative works. Studying these axes jointly lets us test whether engagement, and how predictable it is, varies not only with audience scale but also with account type.

\textbf{Classification per follower base.}
We partitioned the set of collected accounts into three tiers according to the size of their
follower base, adopting the micro/macro/mega taxonomy of Conde and
Casais~\cite{conde2023micro} (also used in the recent Instagram-influencer study
of~\cite{cruz2024characterizing}):

\begin{itemize}
	\item \textbf{Micro-influencers}: $1{,}000 \leq \#followers < 100{,}000$: up-and-coming to mid-sized creators, with limited-to-moderate reach;
	\item \textbf{Macro-influencers}: $100{,}000 \leq \#followers < 1{,}000{,}000$: established mid-sized accounts with broad reach;
	\item \textbf{Mega-influencers}: $\#followers \geq 1{,}000{,}000$: celebrity-scale accounts with a massive follower base.
\end{itemize}

Among the $302{,}392$ \texttt{PERSON} accounts with at least $1{,}000$ followers, \textbf{$203{,}447$ are micro}, \textbf{$86{,}611$ macro}, and \textbf{$12{,}334$ mega} (a further $99{,}687$ fall below the $1{,}000$-follower floor and are excluded from the tier analysis).

\textbf{Classification per entity.} We classified account names in two passes. First, we used the \texttt{spaCy} NLP library~\cite{spacy_2020} to assign each account $u$ an $entity(u)$ $\in\{$\texttt{PERSON}, \texttt{ORG}, \texttt{PRODUCT}, \texttt{GPE} (geo-political entity, e.g.\ a country, city, or state), \texttt{LOC}, \texttt{NORP} (nationalities, religious or political groups), \texttt{FAC}, \texttt{EVENT}, \textsc{work\_of\_art} (\textsc{woa}, titles of creative works such as books, songs, or films)$\}$, processed in 25K-account chunks via \texttt{DuckDB}. Accounts \texttt{spaCy} could not confidently label were marked \texttt{UNKNOWN}. Since high-impact accounts (by followers/engagement) were disproportionately mislabeled or left \texttt{UNKNOWN}---often due to non-canonical handles mixing brand and personal names---we applied a second pass using \texttt{Claude Opus~4.6} as a zero-shot classifier over account name and handle, which takes priority over \texttt{spaCy} labels when available.

Table \ref{tab:entities} lists the final \texttt{entity} labels, which were consolidated into a 6-way split, after excluding \texttt{UNKNOWN} and \textsc{misc} accounts at this level (post counts in parentheses).

\begin{table}[t]
    \centering
    \caption{List of \texttt{entities}.}
    \label{tab:entities}
    \small
    \begin{tabular}{|l|c|r|}
        \hline
        \textbf{Entity label} & \textbf{Subsets} & \textbf{Size} \\ \hline
        \texttt{PERSON (PER)}  & \texttt{PERSON} & 8{,}255{,}603 posts \\ \hline
		\texttt{ORG\_GEO (OG)} & \texttt{ORG} $\cup$ \texttt{NORP} $\cup$ \texttt{GPE} & 7{,}091{,}177 posts\\ \hline
		\texttt{PRODUCTS (PRD)} & \texttt{PRODUCT} & 100{,}818 posts\\ \hline
		\texttt{WORK\_OF\_ART (WoA)} & \texttt{WORK\_OF\_ART} & 39{,}285 posts \\ \hline
		\texttt{EVENT (EVT)} & \texttt{EVENT} & 52{,}007 posts \\ \hline
		\texttt{FAC\_LOC (FL)} & \texttt{FAC} $\cup$ \texttt{LOC} & 150{,}227 posts
        \\ \hline
    \end{tabular}
\end{table}


\begin{figure}[h]
	\centering
	\includegraphics[width=0.9\columnwidth]{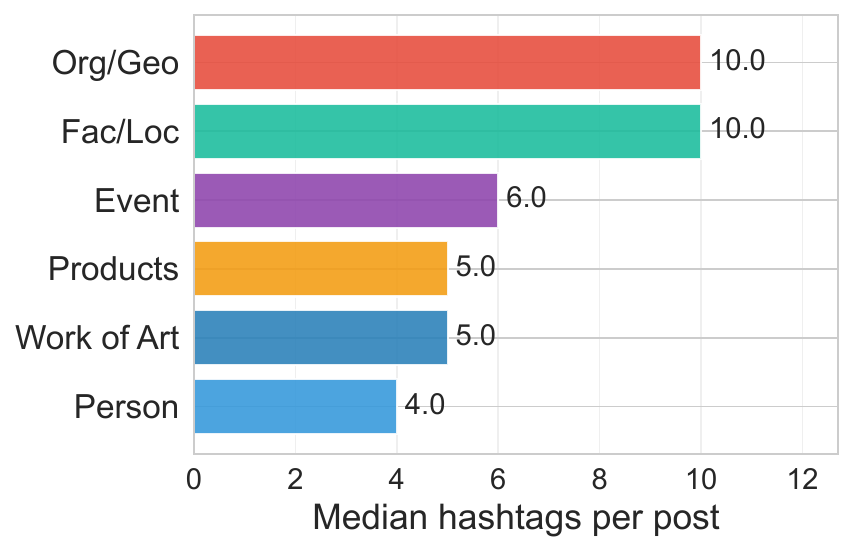}
	\caption{Median number of hashtags per post, by entity.}
	\label{fig:bar_hashtags_merged}
\end{figure}

\subsection{Hashtags and Topics}

Figure \ref{fig:bar_hashtags_merged} compares the median number of hashtags per post across different entity types. Notably, personal accounts (\texttt{PERSON}) use the fewest hashtags, whereas organizational accounts (\texttt{ORG\_GEO}) include the highest number.


We model topics from hashtag co-occurrence rather than full text. Of $\approx$3M distinct hashtags, $\sim$70\% appear for only a single user; we retained the $\approx$7{,}047 hashtags used by at least 250 unique users. Each post's hashtag set was treated as a ``document'' and fed to Gensim's \texttt{LdaMulticore}~\cite{gensim}. We trained the LDA model at $K{=}12$, then labelled each topic manually (e.g., ``Fashion/Beauty,'' ``Fitness/Motivation'') based on its highest-weight hashtags, and finally grouped similar topics together into 8 topics, listed in Table~\ref{tab:glossario_refinado}.

\begin{table}[t]
    \centering
    \caption{List of topics.}
    \label{tab:glossario_refinado}
    \small
    \begin{tabular}{|l|c|l|}
        \hline
        \textbf{Topic} &  & \textbf{Description} \\ \hline
        Fashion        & \texttt{FS}  & Fashion, style, outfits, trends. \\ \hline
        Food           & \texttt{FD}  & Food, cooking, gastronomy, restaurants. \\ \hline
        Fitness/Beauty & \texttt{FIT} & Fitness, gym, beauty, personal care, diet. \\ \hline
        Travel/Geo     & \texttt{TR}  & Travel, tourism, geography, specific places. \\ \hline
        Music/News     & \texttt{MU}  & Music, entertainment, and news. \\ \hline
        Shopping       & \texttt{SH}  & Sales, promotions, digital marketing. \\ \hline
        Lifestyle      & \texttt{LF}  & Family, home, decor, daily life. \\ \hline
        Misc           & \texttt{MC}  & Generic or unclear content. \\
        \hline
    \end{tabular}
\end{table}

\subsection{Metrics and features}
\label{subsec:metrics}

Let $\mathcal{P}$ be the set of all posts, $\mathcal{U}$ the set of accounts, and $\mathcal{H}$ the set of post-filtering hashtags. For each post $p \in \mathcal{P}$, we denote by $\mathcal{L}(p)$ the number of likes and by $\mathcal{C}(p)$ the number of comments; the number of interactions with a post is:
\begin{equation}
\mathcal{I}(p) = \mathcal{L}(p) + \mathcal{C}(p),\quad\forall p\in \mathcal{P}.
\label{eq:interactions}
\end{equation}

We further define the engagement with a post as the number of interactions, normalized by the account's follower base:
\begin{equation}
\textit{eng}(p) = \frac{\mathcal{I}(p)}{\textit{followers}(u)},
\label{eq:eng_rate}
\end{equation}

\noindent where $u$ is the author of $p$ and $\textit{followers}(u)$ denotes the maximum follower count observed across any post by $u$.

For the purpose of analyses, we aggregate the above metrics by \texttt{entity} and by topic.

We denote by $\textit{entity}(u)$ the entity  (from the set of entities listed in Table \ref{tab:entities}) of the account $u\in \mathcal{U}$, by $\textit{hashtags}(p)$, the set of hashtags ($h\in\mathcal{H}$) in post $p\in \mathcal{P}$, and by $\textit{topics}(p)$ the set of topics (listed in Table~\ref{tab:glossario_refinado}) containing at least one hashtag from $\textit{hashtags}(p)$. Finally, we define:
\begin{itemize}
	\item $|\textit{topics}(p)|$: number of topics assigned to a post $p\in\mathcal{P}$;
	\item $\textit{posts}(t) = \{ p \in \mathcal{P} : t \in \textit{topics}(p) \}$: set of posts associated with topic $t$;
	\item $\overline{\mathcal{I}}(t) = \frac{1}{|\textit{posts}(t)|}\sum_{p\in \textit{posts}(t)} \mathcal{I}(p)$: mean number of interactions with a post in $posts(t)$.
\end{itemize}

\textbf{Data processing.}
Given the corpus size ($\approx$15.7M posts post-filtering), all processing uses a zero-memory strategy: data are stored as \texttt{Parquet} and queried via \texttt{DuckDB} without full in-memory loading; iterative steps (e.g., \texttt{spaCy} classification) run in checkpointed chunks; memory/swap usage is monitored throughout via \texttt{psutil} \cite{rodola2009psutil}.


\subsection{Limitations}
Reliance on \texttt{CrowdTangle} imposes temporal and completeness constraints inherent to third-party data brokers, and its 2024 shutdown~\cite{meta2024crowdtangle} caps the dataset with no path to longitudinal extension. Hashtag-based sponsorship detection misses Instagram's native \textit{paid partnership} labels and brand-mention-only disclosures, so any sponsorship proportion should be read as a lower bound. Automated account classification (\texttt{spaCy} + \texttt{Claude}) inevitably yields some false positives/negatives, particularly for accounts representing multiple entity types simultaneously; such ambiguous cases are excluded from fine-grained analyses.
\section{Dataset Characterization}
\label{sec:data_characterization}

We characterize the corpus across the six \texttt{entities} listed in Table \ref{tab:entities}, analyzing how the interaction with sponsored content and the follower base differ across the types of accounts that produce it. 

\begin{figure*}[t]
	\centering
	\includegraphics[width=0.48\textwidth]{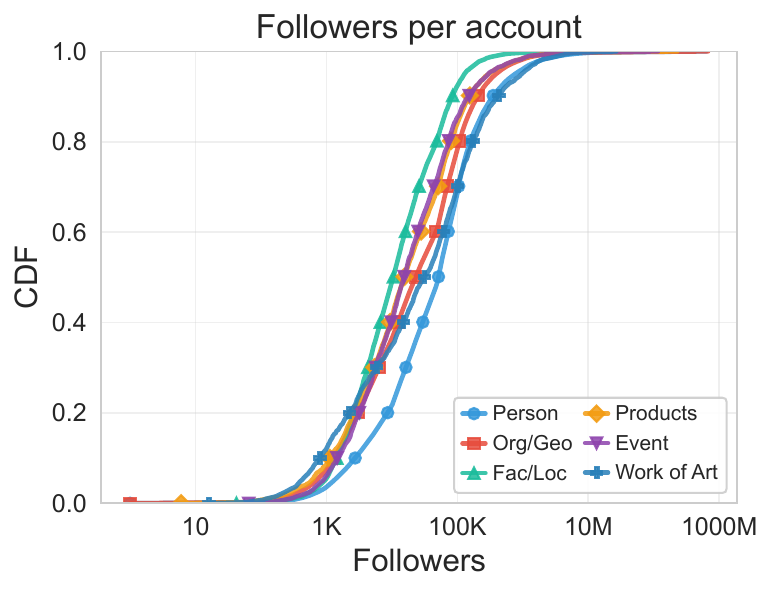}\hfill
	\includegraphics[width=0.48\textwidth]{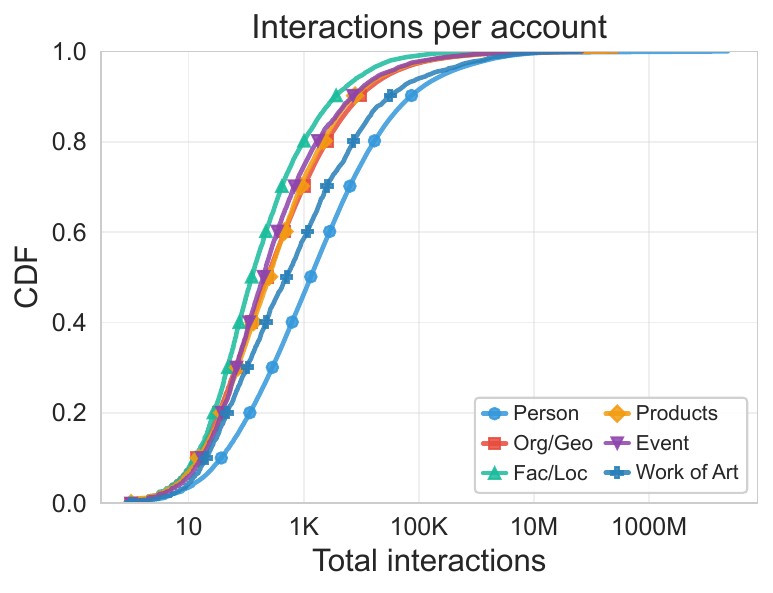}\\[4pt]
	\includegraphics[width=0.48\textwidth]{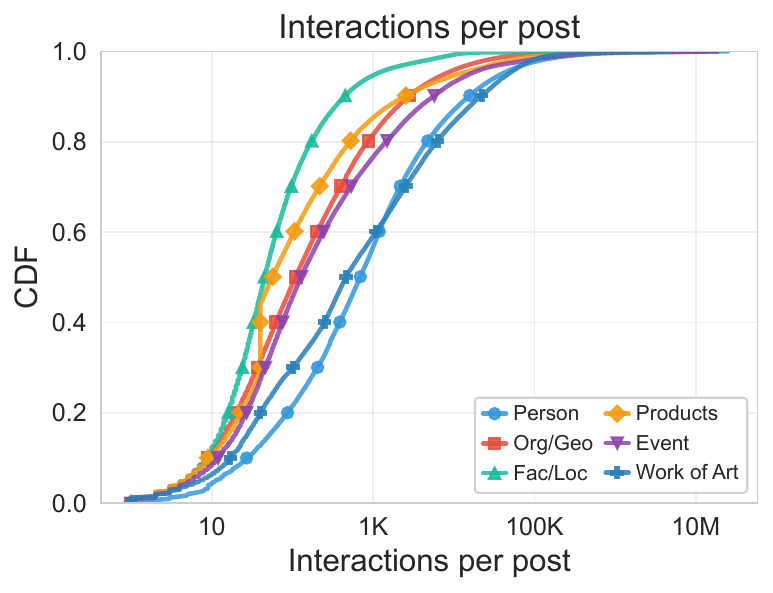}\hfill
	\includegraphics[width=0.48\textwidth]{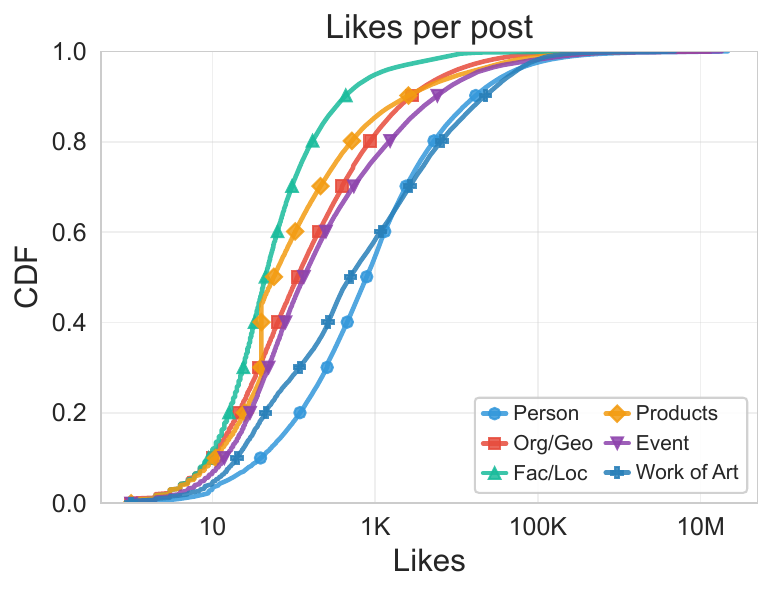}
	\caption{ECDFs (UL)
	followers per account, (UR) interactions per account, (BL) interactions per post,
	and (BR) likes per post.}
	\label{fig:cdf_dataset}
\end{figure*}

\textbf{Followers and Interactions}
Figure~\ref{fig:cdf_dataset} illustrates the (Empirical) Cumulative Distribution Function (CDF) of four parameters: number of followers per account (Upper Left plot), number of interactions per account (Upper Right plot), number of interactions per post (Bottom Left plot), and number of likes per post (Bottom Right plot).
All six \texttt{entities} are strongly heavy-tailed: a small fraction of accounts and posts concentrates most of the engagement; what distinguishes the \texttt{entities} is not the shape of the tail but the level at which it operates. \texttt{PERSON} accounts dominate all four panels (the light-blue curve is the right-most in plots): they reach the highest median follower base ($\approx$51K) and the highest interaction numbers (median $\approx$680 interactions and $\approx$845 likes per post). The \texttt{WORK\_OF\_ART} entity is the closest follower ($\approx$807 median likes per post)---reflecting the strong pull of individuals and creative titles. 
At the quiet end (leftmost curves), \texttt{FAC\_LOC} (green) and \texttt{PRODUCTS}, barely interact ($\approx$40--41 median likes per post), an order of magnitude below \texttt{PERSON}, with \texttt{EVENT} ($\approx$158) and \texttt{ORG\_GEO} ($\approx$117) in between. This ordering---people and creative works engaging far more than products and places---holds across both the per-account and per-post interactions, showing that the gap stems from how each entity is followed and received rather than from posting volume alone.

\begin{figure*}[t]
	\centering
	\includegraphics[width=\textwidth]{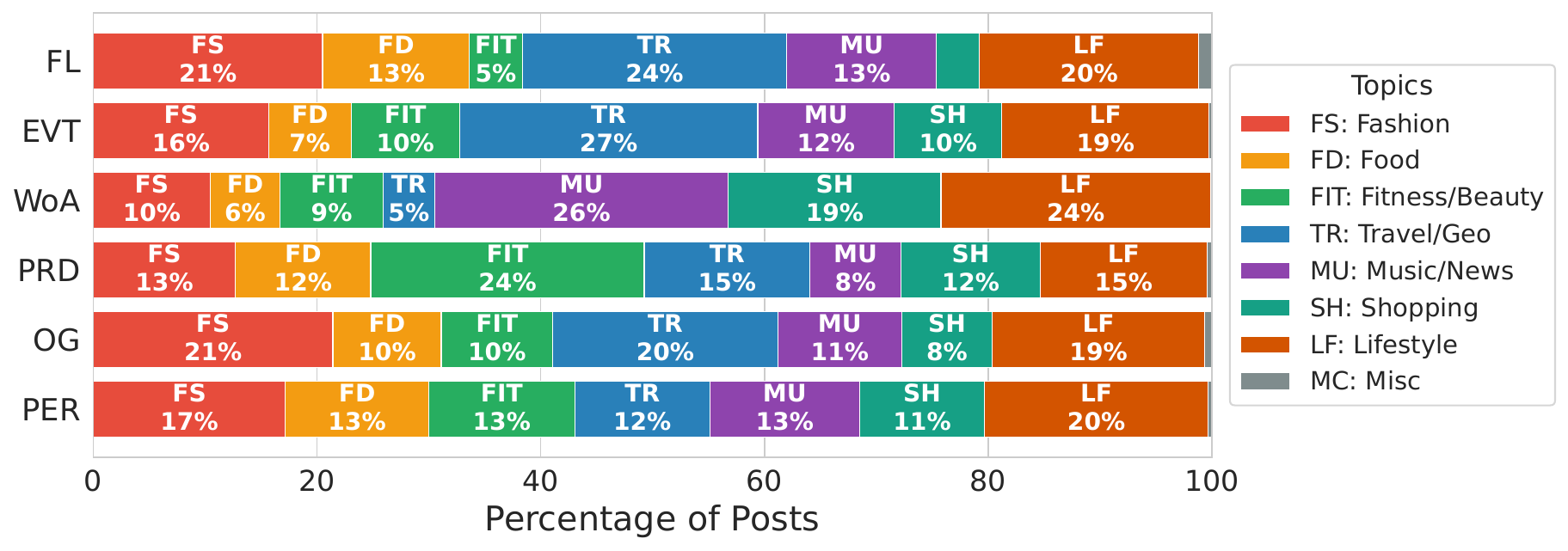}
	\caption{ Topic distribution (\%) per merged entity category (English posts). Each bar sums to 100\%; topic codes follow Table~\ref{tab:glossario_refinado}.}
	\label{fig:stacked_topics_merged_en}
\end{figure*}

\textbf{Topics and Entities.}
To understand the thematic profile of each entity, we fix the entity and analyze the proportional distribution of topics within it. Figure~\ref{fig:stacked_topics_merged_en} shows, as horizontal stacked bars, the thematic composition of each of the six entities (listed in Table \ref{tab:entities}). 
Each color corresponds to a topic listed in Table \ref{tab:glossario_refinado};
inline labels are shown only for slices above 5\% of the total, with smaller slices identified via the side legend. We can see that all entities produce content in all topics in a fairly balanced proportion. Non-surprisingly, the \texttt{WORK\_OF\_ART (WoA)} entity generates a higher relative volume of content within the \texttt{Music/News} topic, while the \texttt{EVENTS (EVT)} entity generates a higher relative volume of content within the \texttt{Travel/Geo} topic. 

\section{Engagement Prediction}
\label{sec:results}

We frame engagement prediction as a \emph{regression} problem: given a sponsored post, how much of its engagement can be explained by the features of the post itself---its content, form, and timing---\emph{without} leveraging the account's follower count or interaction history. Our guiding hypothesis is that \emph{the same posting strategy yields divergent engagement outcomes depending on the underlying account type}. A caption, posting time, or topic that reliably drives interactions for one entity type (e.g., a \texttt{PRODUCT} or an \texttt{EVENT}) may not do so for an individual \texttt{PERSON}. If this hypothesis holds, the entity taxonomy introduced in Section~\ref{sec:data_engineering} is not merely descriptive but \emph{predictive}, effectively separating a \emph{content}-driven view of engagement 
from a \emph{profile}-driven one.
We organize this evaluation around three primary research questions:
\begin{itemize}
  \item How predictable is per-post engagement (likes, comments, and engagement rate) from content, form, and timing alone, measured via variance explained ($R^2$) without using follower counts as a feature?
  \item Does fitting a dedicated model per entity type improve predictive performance, and for which categories? Furthermore, does predictability within the \texttt{PERSON} category vary across the micro, macro, and mega follower tiers defined in Section~\ref{sec:data_engineering}?
  \item Which of the five feature blocks---Content, Form, Time, Topic, and Entity (defined below)---carry the strongest predictive signal, and does their relative importance shift across different entity types?
\end{itemize}

\subsection{Methodology}

To isolate the factors influencing the performance of a sponsored post, we model engagement as a regression task, predicting per-post interaction counts from the features defined in Section~\ref{subsec:metrics}. Crucially, we exclude the account's follower count and historical performance from the feature set. This constraint forces the models to explain engagement based strictly on the characteristics of the \emph{post}, rather than simply recovering the baseline fact that larger accounts naturally accumulate more interactions.

\textbf{Targets.} We independently predict three distinct metrics: the number of likes $\mathcal{L}(p)$, the number of comments $\mathcal{C}(p)$, and the engagement rate $\textit{eng}(p)=\mathcal{I}(p)/\textit{followers}(u)$ as defined in Eq.~\ref{eq:eng_rate}. Likes and comments are modeled separately because they capture different forms of engagement — passive approval vs. active response — and pooling them would obscure this distinction. 
Given the extreme skewness of the raw interaction metrics, all three metrics are target-transformed using a log scale ($\log(y+1)$) prior to training. 
The follower count enters the pipeline \emph{only} to normalize the engagement rate target and is \emph{never} utilized as a predictive feature.

\textbf{Feature blocks.} Each post is represented by five disjoint feature blocks, which also serve as the ablation units for our feature analysis:
\begin{itemize}
  \item \textbf{\texttt{Content}} (\texttt{embed}): A dense semantic embedding of the caption and hashtags generated by the multilingual \texttt{e5-large} sentence encoder~\cite{wang2024multilingual,reimers2019sentence} (1024 dimensions), reduced to 200 dimensions via Principal Component Analysis (PCA);
  \item \textbf{\texttt{Form}}: Caption length, hashtag count, and number of account mentions;
  \item \textbf{\texttt{Time}}: Day of the week, month, year, local hour, and a weekend indicator flag;
  \item \textbf{\texttt{Topic}}: The topic indicators and overall topic count;
  \item \textbf{\texttt{Entity}}: Entity category and language (one-hot encoded).
\end{itemize}
The \texttt{Content} block represents our key methodological departure from classic hashtag-and-topic analyses. Rather than compressing a caption into coarse LDA buckets, the dense embedding preserves granular \emph{within-topic} variation that, as we demonstrate, carries the vast majority of the predictive signal.

\textbf{Prediction algorithm.} We evaluated two tree-based ensembles: Random Forests and Gradient Boosted Trees (\texttt{xgboost}~\cite{chen2016xgboost}). Because \texttt{xgboost} (with hyperparameters tuned via a held-out validation split) consistently outperformed Random Forests by a stable margin, it is the sole model reported in our results. 

We measure predictive performance using the coefficient of determination ($R^2$) on a held-out test set:
\begin{equation}
R^2 = 1 - \frac{\sum_i \left(y_i-\hat{y}_i\right)^2}{\sum_i \left(y_i-\bar{y}\right)^2},
\label{eq:r2}
\end{equation}
where $y_i$ represents the observed ground-truth value, $\hat{y}_i$ is the model prediction, and $\bar{y}$ is the test-set mean; an $R^2{=}1$ denotes perfect prediction, while $R^2{=}0$ indicates performance equivalent to a constant-mean baseline. Models are trained using an 80/20 post-level random split. We note that splitting by post (rather than by account) allows posts from the same profile to appear across both training and test folds. Consequently, the absolute $R^2$ values reported here represent an upper bound on out-of-account generalization.


\textbf{Sampling.} For each analytical segment, we extract a reproducible random sample of up to 50{,}000 posts using deterministic \texttt{DuckDB} \texttt{bernoulli} sampling (seed~42), preserving the real-world, heavy-tailed interaction distribution (Section~\ref{sec:data_characterization}); segments below this threshold contribute all available posts. We deliberately evaluate our models on this natural, skewed distribution to ensure that the reported metrics accurately reflect performance on the authentic mix of posts a model would encounter in production, rather than on an artificially re-weighted dataset.

  \begin{table*}[t]
    \centering
    \small
    \setlength{\tabcolsep}{5pt}
    \caption{Per-entity engagement-prediction performance.}
    \label{tab:metrics}
    \begin{tabular}{l ccc ccc ccc}
      \toprule
       & \multicolumn{3}{c}{\textbf{Likes}} & \multicolumn{3}{c}{\textbf{Comments}} & \multicolumn{3}{c}{\textbf{Engagement}} \\
      \cmidrule(lr){2-4}\cmidrule(lr){5-7}\cmidrule(lr){8-10}
      \textbf{Entity} & $R^2$ & MAE & RMSE & $R^2$ & MAE & RMSE & $R^2$ & MAE & RMSE \\
      \midrule
      \texttt{ALL} (baseline)  & 0.40 & 1.60 & 2.22 & 0.49 & 1.04 & 1.37 & 0.42 & 1.14 & 1.60 \\
      \texttt{PERSON}          & 0.33 & 1.70 & 2.48 & 0.43 & 0.99 & 1.37 & 0.39 & 1.08 & 1.56 \\
      \texttt{ORG\_GEO}        & 0.44 & 1.39 & 1.89 & 0.39 & 1.02 & 1.35 & 0.39 & 1.14 & 1.57 \\
      \texttt{PRODUCTS}        & 0.61 & 0.93 & 1.51 & 0.66 & 0.70 & 1.08 & 0.58 & 0.75 & 1.23 \\
      \texttt{FAC\_LOC}        & 0.58 & 0.86 & 1.25 & 0.53 & 0.68 & 0.94 & 0.59 & 0.75 & 1.06 \\
      \texttt{EVENT}           & 0.74 & 0.98 & 1.44 & 0.66 & 0.85 & 1.15 & 0.58 & 0.84 & 1.19 \\
      \texttt{WORK\_OF\_ART}   & 0.74 & 0.95 & 1.46 & 0.64 & 0.85 & 1.19 & 0.63 & 0.76 & 1.13 \\
      \bottomrule
    \end{tabular}
  \end{table*}

\begin{figure*}[t]
  \centering
  \includegraphics[width=0.32\textwidth]{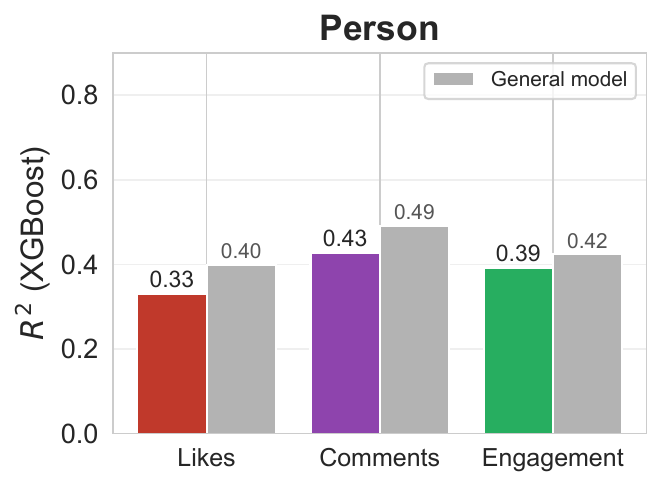}\hfill
  \includegraphics[width=0.32\textwidth]{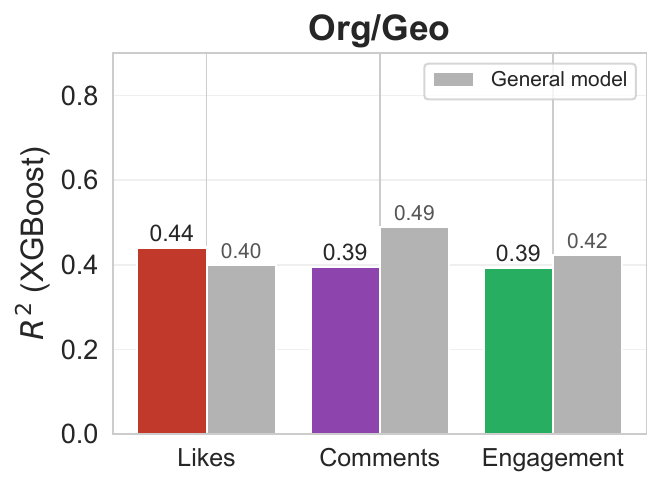}\hfill
  \includegraphics[width=0.32\textwidth]{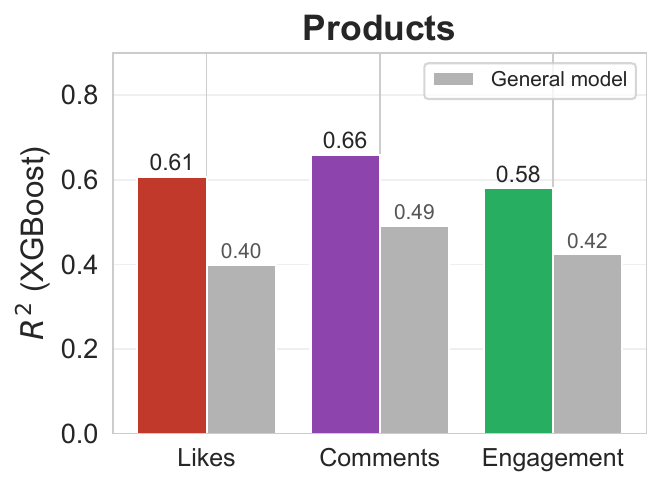}\\[2pt]
  \includegraphics[width=0.32\textwidth]{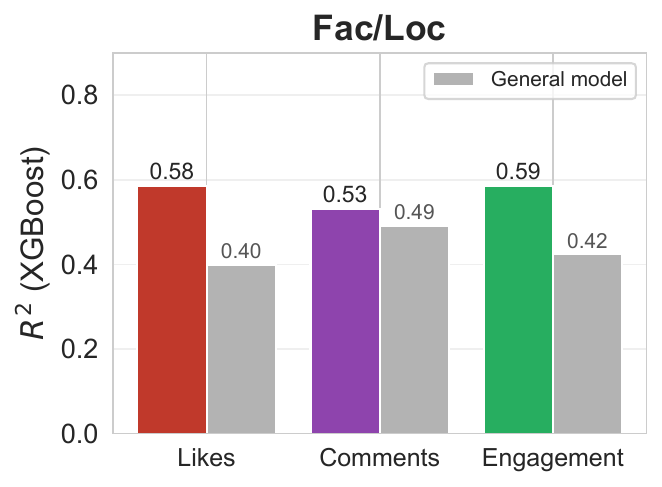}\hfill
  \includegraphics[width=0.32\textwidth]{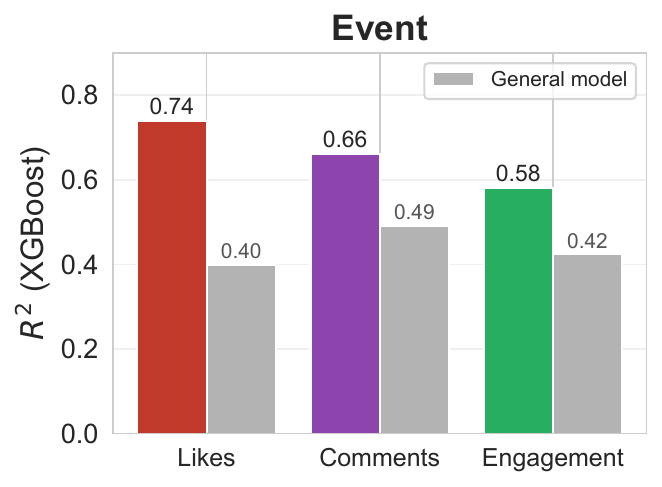}\hfill
  \includegraphics[width=0.32\textwidth]{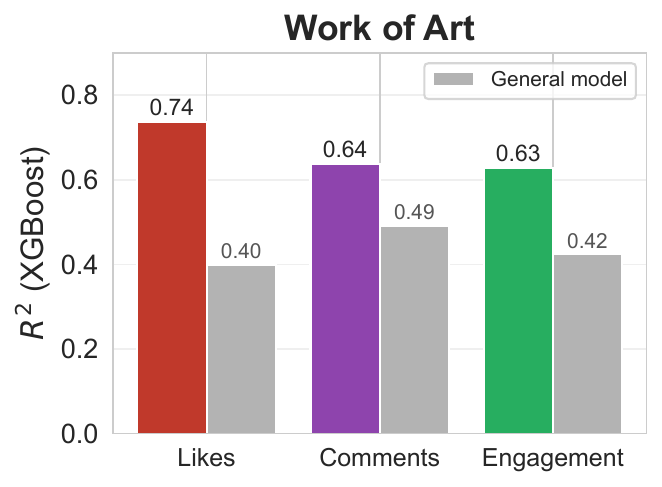}
  \caption{Interaction and engagement-prediction $R^2$ (XGBoost). In each panel, the gray bar is the baseline (\texttt{ALL}) model
  for the same target; colored bars indicate a dedicated per-entity model's performance.}
  \label{fig:r2_natural}
\end{figure*}

\subsection{Results}

\textbf{Splitting by entity.} Figure~\ref{fig:r2_natural} details the $R^2$ scores obtained per segment using \texttt{xgboost}. A single pooled (\texttt{ALL}) baseline model explains $R^2{=}0.40$ of the variance in likes, $0.49$ in comments, and $0.42$ in engagement---a highly substantial share given the complete absence of follower metadata. Crucially, fitting a dedicated model \emph{per entity type} sharply increases predictability for brand-centric and niche categories: \texttt{PRODUCTS} (0.61/0.66/0.58), \texttt{FAC\_LOC} (0.58/0.53/0.59), \texttt{EVENT} (0.74/0.66/0.58), and \texttt{WORK\_OF\_ART} (0.74/0.64/0.63) all outperform the general baseline model by a wide margin. Conversely, the dedicated \texttt{PERSON} model (0.33/0.43/0.39) closely tracks the \texttt{ALL} baseline, which is expected given that individual creators thoroughly dominate the underlying corpus. Table~\ref{tab:metrics} provides the comprehensive evaluation suite, pairing each $R^2$ metric with its corresponding Mean Absolute Error (MAE) and Root Mean Squared Error (RMSE), calculated on the held-out test set.

\begin{figure}[h]
  \centering
  \includegraphics[width=\columnwidth]{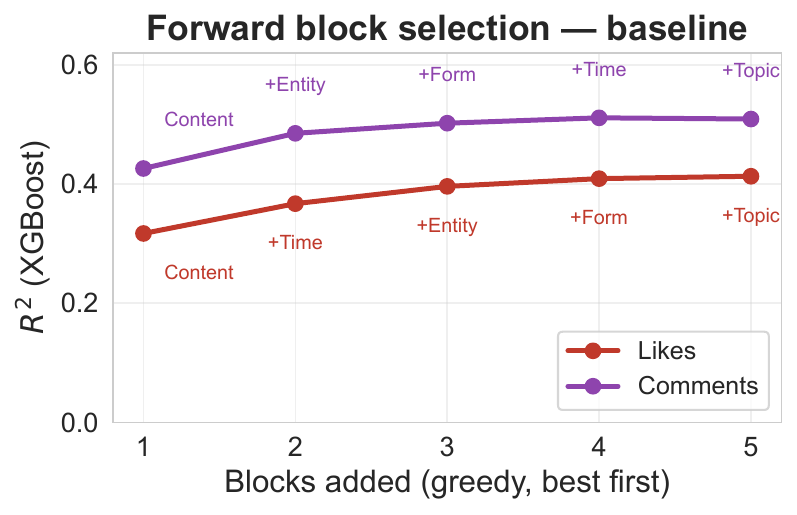}
  \caption{Forward block selection on the baseline (\texttt{ALL}) model.}
  \label{fig:ablation}
\end{figure}

\textbf{Constructive Ablation.} Figure~\ref{fig:ablation} illustrates a forward block selection execution on the general model. The dense \texttt{Content} embedding is selected first for both prediction targets, single-handedly capturing $R^2{=}0.32$ for likes and $0.43$ for comments; subsequent feature blocks yield diminishing returns. Notably, the second feature block selected is \texttt{Time} for likes but \texttt{Entity} for comments, confirming that institutional entity type exerts a more profound influence on the conversational mechanics of commenting than on the lower-friction behavior of liking. The (LDA) \texttt{Topic} block is selected last and offers negligible predictive gains, indicating that our dense semantic embedding already subsumes the necessary topical signals. A complementary leave-one-out validation check confirms this pattern: removing \texttt{Content} induces the most severe drop in performance ($\Delta R^2\!\approx\!0.12$--$0.13$), whereas omitting \texttt{Topic} incurs no measurable cost. This stands as a core methodological takeaway: a dense semantic representation of the caption serves as the primary engine of the predictive model, with entity categorization acting as its most effective structural complement.

The per-post engagement distributions (visualized in the right-hand panels of Figure~\ref{fig:cdf_dataset}) make this operational difference concrete: \texttt{PERSON} and \texttt{WORK\_OF\_ART} posts secure the highest median engagement, whereas \texttt{PRODUCTS} and \texttt{FAC\_LOC} profiles operate under much quieter interaction bounds. These distinct operating baselines highlight why a single pooled model remains suboptimal, and validate the predictive lift gained from an entity-segmented architecture.

\begin{figure}[h]
  \centering
  \includegraphics[width=\columnwidth]{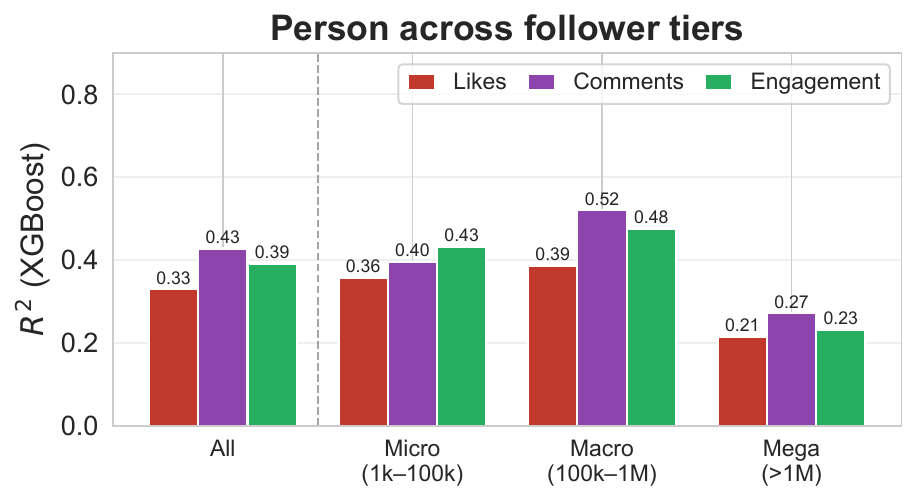}
  \caption{Splitting \texttt{PERSON} accounts by follower base.}
  \label{fig:person_tiers}
\end{figure}

\textbf{Splitting by follower base.} Finally, we evaluate whether engagement predictability varies within a single entity category as a function of audience size. Restricting our data to the \texttt{PERSON} category, we fit independent models across the three follower tiers proposed by Conde and Casais~\cite{conde2023micro}: \emph{micro} (1K--100K), \emph{macro} (100K--1M), and \emph{mega} ($>$1M) followers. Figure~\ref{fig:person_tiers} tracks the resulting $R^2$ scores alongside the pooled \texttt{PERSON} baseline (\texttt{ALL}); the absolute follower metric is kept completely excluded from the features. Predictability is highly non-uniform across these tiers: it peaks distinctively among \emph{macro}-influencers (e.g., reaching a comments $R^2$ of $0.52$ versus the baseline $0.43$) and hits its lowest point among \emph{mega} accounts (likes/comments/engagement at $0.21/0.27/0.23$). Structurally, the engagement patterns of massive, mainstream celebrities are the least explainable via localized content, form, and timing attributes—implying that their reach is governed by external cultural forces outside the immediate post—whereas mid-tier (\emph{macro}) creators represent the most highly content-predictable cohort on the platform.

\section{Conclusion}
\label{sec:conclusion}


We studied per-post engagement with sponsored content on Instagram as a
regression problem, predicting likes, comments, and engagement rate from the
post itself---content, form, and timing.
Using interpretable, tree-based models
(\textsc{xgboost}~\cite{chen2016xgboost}) rather than deep architectures, we find
that a meaningful share of engagement is predictable from the post alone: a
pooled model reaches $R^2{=}0.40$ for likes and $R^2{=}0.49$ for comments.
Crucially, this predictability is far from uniform. Splitting accounts by entity type shows that brand-like and niche categories---\texttt{PRODUCTS},
\texttt{EVENT}, \textsc{WORK\_OF\_ART}, and \texttt{FAC\_LOC}---are substantially
more predictable than the pooled baseline (up to $R^2{=}0.74$), whereas
\texttt{PERSON} accounts track the pooled baseline. A feature-block ablation
explains why: the dense semantic representation of the caption (the \texttt{e5}
sentence embedding~\cite{wang2024multilingual,reimers2019sentence}) dominates,
far outweighing LDA topics, while entity type is the strongest structural
complement, especially for comments. Within \texttt{PERSON}, predictability is
highest for macro influencers and lowest for mega-celebrities, whose engagement
is least explainable from content, form, and timing alone.

Taken together, these results carry a single message: \emph{one size does not
fit all}. The mechanisms that drive engagement, and the degree to which
engagement can be anticipated at all, vary sharply with the type of account
behind a sponsored post, so a single pooled model both under-fits niche
categories and obscures these differences. For practitioners, the finding that
\emph{what} and \emph{how} an account posts is predictive independently of
audience size suggests a lever that even smaller accounts can exploit; for
researchers, it argues that account taxonomy should be a first-class variable
when modeling social-media engagement, and that interpretable models suffice to
expose these structural patterns.

\textbf{Limitations.}
Our findings should be read with several caveats. Sponsorship is detected from
disclosure hashtags, which yields a high-precision but conservative lower bound
on sponsored activity, since undisclosed paid content is missed. The corpus is
drawn from \texttt{CrowdTangle}, whose coverage is bounded in time and which Meta
discontinued in 2024~\cite{meta2024crowdtangle}, so newer dynamics are not
captured. Account-type assignment is automated and therefore imperfect, and our
analysis is correlational rather than causal: we characterize what predicts
engagement, not what causes it. Finally, the 80/20 split is at the post level, so
different posts from the same account can fall in both folds, which may inflate
predictability for accounts with many posts.

\textbf{Future Work}
A natural next step is a cross-platform comparison, contrasting these entity-type
patterns with sponsored engagement on TikTok, YouTube, and Facebook. The
correlational findings also motivate causal designs---e.g., quasi-experimental or
matched analyses---to isolate the effect of content choices on engagement.


\begin{acks}
CNPq, FAPEMIG, CAPES.
\end{acks}

\bibliographystyle{ACM-Reference-Format}
\bibliography{references_sponsored}


\end{document}